\documentclass[aps,prl,reprint,superscriptaddress]{revtex4-1}
\usepackage{graphicx}

\begin{document}


\title{Ultrafast x-ray diffraction of a ferroelectric soft mode driven by broadband \protect\\terahertz pulses}
\author{S. Gr\"ubel}\email{sebastian.gruebel@psi.ch}\affiliation{Swiss Light Source, Paul Scherrer Institut, 5232 Villigen PSI, Switzerland.}\affiliation{Institute for Quantum Electronics, Eidgen\"ossische Technische Hochschule (ETH) Z\"urich, Auguste-Piccard-Hof 1, 8093 Z\"urich, Switzerland.}
\author{J. A. Johnson}\affiliation{Swiss Light Source, Paul Scherrer Institut, 5232 Villigen PSI, Switzerland.}\affiliation{Department of Chemistry and Biochemistry, Brigham Young University (BYU) Provo, UT, USA}
\author{P. Beaud}\affiliation{Swiss Light Source, Paul Scherrer Institut, 5232 Villigen PSI, Switzerland.}\affiliation{SwissFEL, Paul Scherrer Institut, 5232 Villigen PSI, Switzerland}
\author{C. Dornes}\affiliation{Institute for Quantum Electronics, Eidgen\"ossische Technische Hochschule (ETH) Z\"urich, Auguste-Piccard-Hof 1, 8093 Z\"urich, Switzerland.}
\author{A. Ferrer}\affiliation{Swiss Light Source, Paul Scherrer Institut, 5232 Villigen PSI, Switzerland.}\affiliation{Institute for Quantum Electronics, Eidgen\"ossische Technische Hochschule (ETH) Z\"urich, Auguste-Piccard-Hof 1, 8093 Z\"urich, Switzerland.}
\author{V. Haborets}\affiliation{Uzhgorod National University, Institute for Solid State Physics and Chemistry, 3 Narodna Square, 88000 Uzhgorod, Ukraine.}
\author{L. Huber}\affiliation{Institute for Quantum Electronics, Eidgen\"ossische Technische Hochschule (ETH) Z\"urich, Auguste-Piccard-Hof 1, 8093 Z\"urich, Switzerland.}
\author{T. Huber}\affiliation{Institute for Quantum Electronics, Eidgen\"ossische Technische Hochschule (ETH) Z\"urich, Auguste-Piccard-Hof 1, 8093 Z\"urich, Switzerland.}
\author{A. Kohutych}\affiliation{Uzhgorod National University, Institute for Solid State Physics and Chemistry, 3 Narodna Square, 88000 Uzhgorod, Ukraine.}
\author{T. Kubacka}\affiliation{Institute for Quantum Electronics, Eidgen\"ossische Technische Hochschule (ETH) Z\"urich, Auguste-Piccard-Hof 1, 8093 Z\"urich, Switzerland.}
\author{M. Kubli}\affiliation{Swiss Light Source, Paul Scherrer Institut, 5232 Villigen PSI, Switzerland.}\affiliation{Institute for Quantum Electronics, Eidgen\"ossische Technische Hochschule (ETH) Z\"urich, Auguste-Piccard-Hof 1, 8093 Z\"urich, Switzerland.}
\author{S. O. Mariager}\affiliation{Swiss Light Source, Paul Scherrer Institut, 5232 Villigen PSI, Switzerland.}
\author{J. Rittmann}\affiliation{LSU, Ecole Polytechnique Federale de Lausanne, 1015 Lausanne,Switzerland}
\author{J. I. Saari}\affiliation{Institute for Quantum Electronics, Eidgen\"ossische Technische Hochschule (ETH) Z\"urich, Auguste-Piccard-Hof 1, 8093 Z\"urich, Switzerland.}
\author{Y. Vysochanskii}\affiliation{Uzhgorod National University, Institute for Solid State Physics and Chemistry, 3 Narodna Square, 88000 Uzhgorod, Ukraine.}
\author{G. Ingold}\affiliation{Swiss Light Source, Paul Scherrer Institut, 5232 Villigen PSI, Switzerland.}\affiliation{SwissFEL, Paul Scherrer Institut, 5232 Villigen PSI, Switzerland}
\author{S. L. Johnson}\affiliation{Institute for Quantum Electronics, Eidgen\"ossische Technische Hochschule (ETH) Z\"urich, Auguste-Piccard-Hof 1, 8093 Z\"urich, Switzerland.}




\begin{abstract}
Intense, few-cycle pulses in the terahertz frequency range have strong potential for schemes of control over vibrational modes in solid-state materials in the electronic ground-state. Here we report an experiment using single cycle terahertz pulses to directly excite lattice vibrations in the ferroelectric material $\mathrm{Sn_2P_2S_6}$ and ultrafast x-ray diffraction to quantify the resulting structural dynamics. A model of a damped harmonic oscillator driven by the transient electric field of the terahertz pulses describes well the movement of the Sn$^{2+}$ ion along the ferroelectric soft mode. Finally, we describe an anharmonic extension of this model which predicts coherent switching of domains at peak THz-frequency fields of 790 kV/cm. 
\end{abstract}


\maketitle


The ultimate speed of ferroelectric polarization switching is of great interest for potential data storage devices using ferroelectric materials. In these materials, the reversal of the polarization by application of an electric field is normally described as a nucleation process followed by motion of domain walls on the order of several m/s. For typical device dimensions this implies a switching process on a nanosecond timescale \cite{tybellprl89,grigorievprl2006}. A potentially faster approach that could lead to switching on picosecond time scales is to use short electromagnetic pulses to drive vibrational modes coupled to the polarization \cite{fahyprl1994}. In this case the mechanism inducing polarization reversal is very different: the excitation of a large amplitude terahertz (THz) lattice vibration induces a spatially uniform coherent structural motion comparable to the atomic motion in the ferroelectric transition which drives the structure to the opposite polarization state. The excitation of such coherent phonons with eigenfrequencies in the terahertz frequency region can be achieved indirectly using ultrashort optical light pulses through displacive excitation or impulsive stimulated raman scattering (ISRS)\cite{yanjpc1987,zeigerprb1992}. The very high intensities of optical radiation required to drive large-scale vibrations in either of these two cases, however, lead to material damage which limits the potential of these mechanisms for control applications. An alternate solution is to use low-frequency electromagnetic radiation to drive infrared active modes directly with the oscillations of the electric field. Recent models of dynamics in simple ferroelectric materials have indicated that this is possible with intense, coherent pulses of radiation with frequencies ranging from 0.5-30 THz and peak fields up to several MV/cm \cite{qiprl2009}. Only recently has it become possible to generate such pulses under specialized conditions in the laboratory \cite{HiroriAPL2011,VicarioJMO2013,ShalabyNC2015}.

In order to test these models against experiment, a quantitative, time-resolved measurement of the structure of the unit cell during and after interaction with a broadband THz pulse would be extremely useful. Because of the frequencies involved, a time resolution of less than one picosecond is needed. Conventionally, ultrafast optical methods are often used to extract information about transient properties of materials and quantitative structural infomation can be inferred~\cite{KatayamaPRL2012}, but these results are themselves highly model dependent. Time-resolved x-ray diffraction offers a more direct measure of quantitative structural motions. Time resolved x-ray diffraction studies of ultrafast structural dynamics driven by ultrashort pulses extending from the visible \cite{ElsaesserJCP2014} to the mid-infrared \cite{FoerstACR2015} range of the optical spectrum have already been demonstrated on several different systems. 
Specifically, Cavalleri et al. have reported on coherent phonon-polariton structural dynamics in $\mathrm{LiTaO_3}$ driven with ISRS \cite{CavalleriNature2006}. The difficulty of generating large-amplitude THz pulses under conditions amenable for a time resolved x-ray diffraction experiment has so far made it challenging to drive a measurable coherent structural motion with externally generated THz pulses.

In this paper we demonstrate a direct measurement of atomic displacements driven by a single cycle THz pulse. A $\mathrm{Sn_2 P_2 S_6}$ bulk ferroelectric crystal was irradiated with broadband THz pulses with frequency range spanning the ferroelectric soft mode resonance; structural changes were monitored using ultrafast x-ray diffraction. We model the diffraction signal and quantify the displacement of the Sn$^{2+}$ ion with respect to its displacement from the paraelectric towards the ferroelectric phase.  

\begin{figure}[h]
\includegraphics[width = 8.6cm]{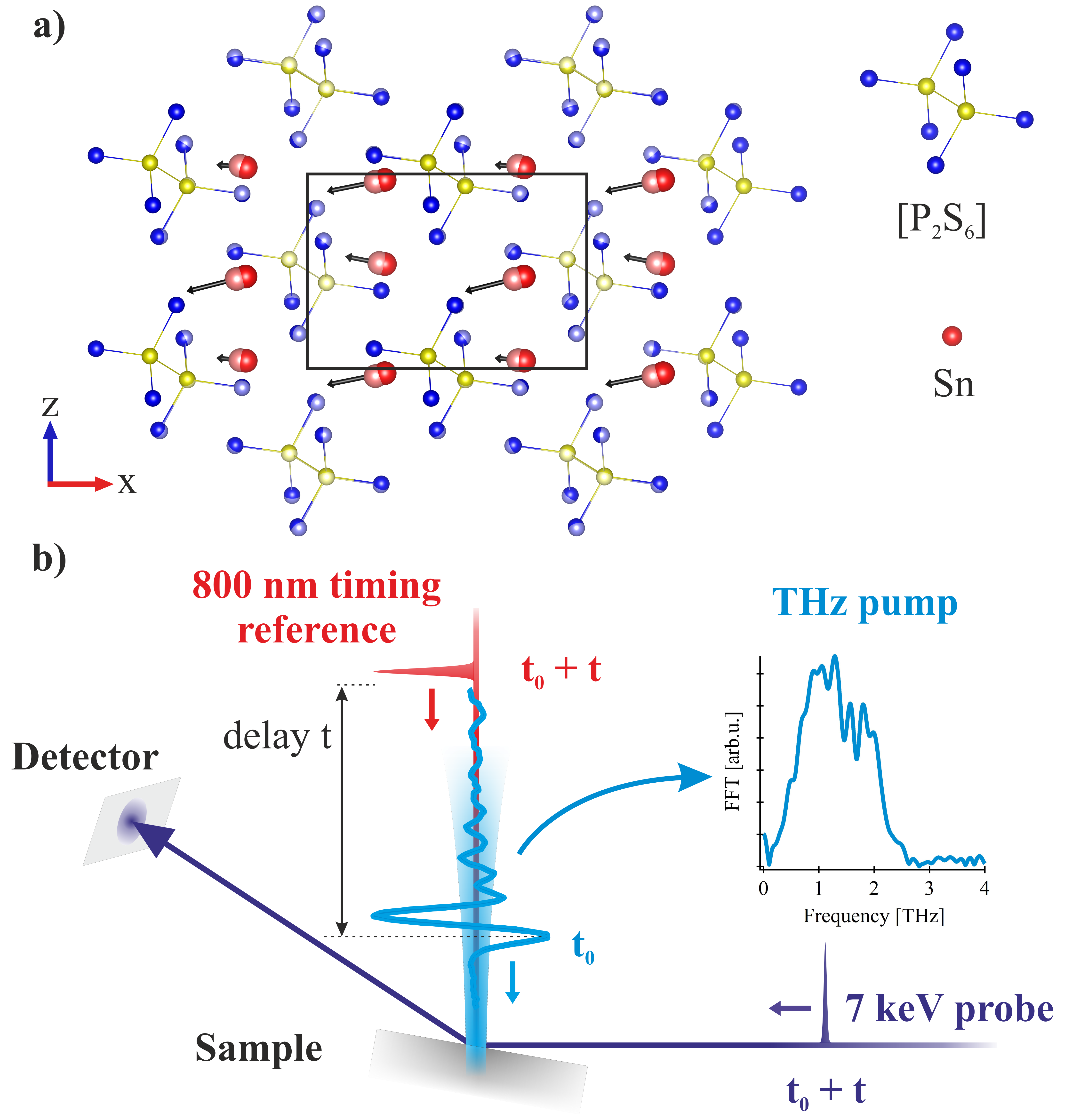}
\caption{\label{l.figure1} a) Overlap of $\mathrm{Sn_2 P_2 S_6}$ crystal structures  in the low (dark) and high (bright) temperature phase. Arrows indicate the direction of the Sn$^{2+}$ displacements to their high temperature positions. b) Experimental scheme. 800 nm pulses are used prior to the measurements to overlap THz and x-ray pulses in time and space on the sample (as described in the text). }
\end{figure}

At temperatures below $\mathrm{T_c\ \approx}$ 337 K, the compound $\mathrm{Sn_2P_2S_6}$ is ferroelectric with space group $\mathrm{Pn}$ and cell parameters $a = 9.378$ \r{A}, $b= 7.488$ \r{A}, $c= 6.513$ \r{A}, $\beta = 91.15^\circ$ \cite{dittmarznatur1974}. The ferroelectricity manifests as a strong spontaneous polarization in [100] direction that correlates mainly to a shift of the Sn$^{2+}$ ions from centro-symmetric positions along the [100] direction with respect to the [P$_2$S$_6$]$^{4-}$ anion complexes (Figure \ref{l.figure1}) \cite{scottjssc1992}. When heated above $\mathrm{T_c}$ the compound undergoes a displacive phase transition and becomes paraelectric with space group $\mathrm{P21/n}$. Experimental data relate the transition to the softening of a terahertz frequency lattice vibration of  symmetry $\mathrm{A^{\prime}}$ responsible of driving the Sn$^{2+}$ ions mainly along the [100] axis \cite{vysochanskiiftt1978,grabarftt1984,eijtepjb1998}.

Pump probe time resolved x-ray diffraction measurements were carried out at the hard x-ray FEMTO slicing source at the Swiss Light Source (SLS) \cite{beaudprl2007}. Femtosecond laser pulses are produced in a conventional Kerr-lens mode-locked oscillator phase-locked to a submultiple of the synchrotron RF-master oscillator and split into pump and probe branches. The pump pulses enter a stretcher and a regenerative Ti:sapphire amplifier (800 nm, 1 kHz, 2.4 mJ) for chirped pulse amplification. A vacuum transfer line transports the uncompressed optical pulses to the experimental hutch. Inside the experimental hutch, the high power pulses enter a pulse compressor with the majority of the light directed into an optical parametric amplifier (OPA) to generate $\sim$100 fs pulses at a wavelength of 1310 nm. A small amount of 800 nm laser light is picked off before the OPA to facilitate temporal and spatial overlap as described below. Intense THz pulses are generated by optical rectification of the infrared pulses in an 2-{3-(4-hydroxystyryl)-5,5-dimethylcyclohex-2-enylidene}malononitrile (OH1) organic crystal \cite{BrunnerOE2008,HunzikerJOSB2008,KwonAFM2008,RuchertOL2012}. Two off-axis parabolic (OAP) mirrors are used to direct and focus the THz pulses onto the sample. The first OAP is used to collect and collimate the emitted terahertz pulses. The second OAP is placed close to the sample and focuses the beam to a small spot size to achieve a large electric field. The smallest spot size measured at the sample position with a THz camera was $390\ \mathrm{\mu m}\times 410\ \mathrm{\mu m}$ full width half maximum (FWHM). For time dependent measurements, the THz pulses were delayed in time with respect to the x-ray pulses by increasing the path length of the uncompressed pulses using a delay stage. 

The generated x-ray probe pulses (2 kHz, 7 keV, $\sim$140 fs FWHM) \cite{beaudprl2007} were incident at a 10$^{\circ}$ grazing angle with respect to the sample surface and at 90$^{\circ}$ angle with respect to the THz pulses (Figure \ref{l.figure1}). A Kirkpatrick-Baez (KB) mirror focused the beam vertically to a size of 10 $\mathrm{\mu m}$ resulting in a $\mathrm{250\ \mu m\ \times\ 60\ \mu m}$ spot size on the sample. The spot size was much smaller than the THz spot size to ensure a homogeneous lateral excitation profile over the probe pulse. The intensity of the diffracted x-ray pulses from the sample was recorded by a gated avalanche photodiode (APD). The experimental time resolution depended predominantly on the duration of the x-ray pulses and on the geometry of the experiment and was estimated to be $\sim$240 fs FWHM.

Given the non-collinear geometry of the experiment, a 2-step procedure was followed to ensure the spatial and temporal overlap of the x-ray and THz pulses on the sample. First, the transient reflectivity of a bismuth (111) or  a tellurium (100) Bragg reflection was measured at varying delay times between high fluence 800 nm pump pulses and the x-ray pulses. The transient grazing incidence x-ray diffraction measurements were used to overlap the laser pulses and the x-ray pulses in space and time on the bismuth sample \cite{JohnsonPRL2008}. In a second step we replaced this crystal by a gallium phosphide (GaP) crystal. We used the electro-optic (EO) effect induced by the THz pulses in the crystal to overlap the THz pulses spatially and temporally with the 800 nm pulses.

\begin{figure}[h]
\includegraphics[width = 8.6cm]{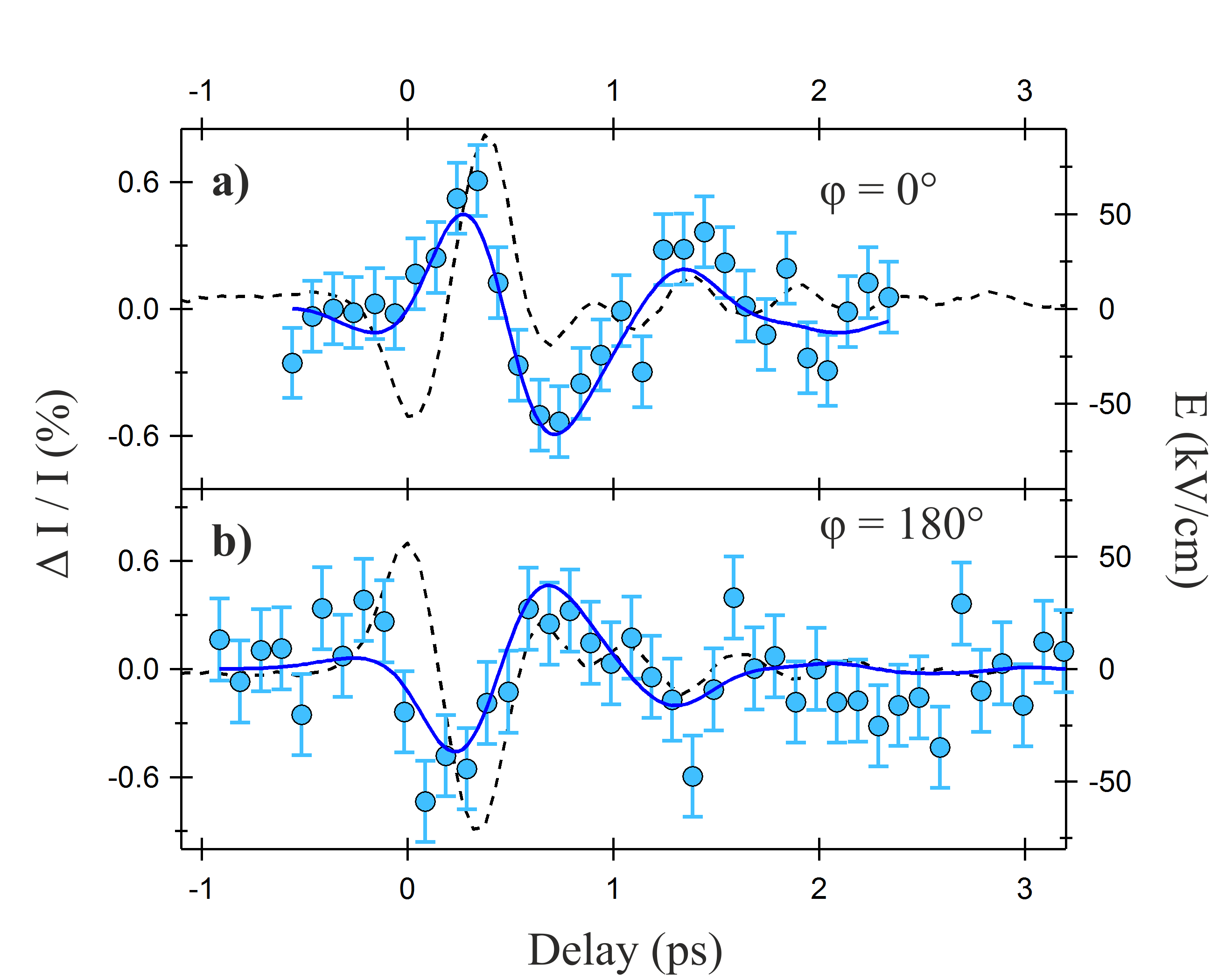}
\caption{\label{l.figure2} (full circle) Time evolution of the normalized peak diffraction intensity of the (332) reflection at room temperature. (full line) Fit using a damped harmonic oscillator model described in the text. (dashed line) EO measurement of the THz pump pulse. The sign of the electric field is reversed between the two measurements a) ($\varphi = 0^{\circ}$) and b)  ($\varphi = 180^{\circ}$).}
\end{figure}

The sample used in the experiment was a single crystal of $\mathrm{Sn_2P_2S_6}$ cut to the (010) surface. The temperature of the crystal was controlled  in the range of 200 K to 300 K by a nitrogen cryojet mounted close to the crystal surface. The maximum terahertz electric field measured at the sample position was 120 kV/cm. The terahertz polarization was set parallel to the [100] axis of the $\mathrm{Sn_2 P_2 S_6}$ crystal. We measured the diffracted intensity $I(t)$ of the (332) Bragg reflection at room temperature (300 K) and at various time delays $t$ between the THz pump and x-ray probe pulses. Figure \ref{l.figure2}(a) shows the transient diffraction intensity ${\Delta I(t)}/{I_0} = ({I(t) - I_0})/{I_0}$ at different time delays $t$, normalized to the equilibrium diffraction intensity $I_0$ and averaged over many pulses. An EO measurement of the driving THz pulse arriving at time $t_0$ was taken prior to each diffraction measurement and is shown as well. In Figure \ref{l.figure2}(b) the electric field of the driving THz pulse has been reversed by rotating the OH1 crystal by 180 degrees, resulting in a sign change of the transient diffraction intensity. 

To model the data, we need to consider the interaction of the electric field of the THz pulses with the vibrational modes that can potentially contribute to the diffraction signal.  The absorption of the THz pulses inside the material creates an inhomogeneous excitation profile along a direction perpendicular to the surface. Consequently, the measured diffraction signal contains contributions from regions with varying excitation levels. To treat this effect, we conceptually divide the sample into thin slices parallel to the surface and consider the interaction of the electric field with the vibrational modes in each slice. 

Raman and neutron scattering together with submillimeter spectroscopy studies of $\mathrm{Sn_2P_2S_6}$ indicate that the soft mode eigenvector and frequency change strongly with temperature, especially near the phase transition \cite{vysochanskiiftt1978,volkovSPSS1983,grabarftt1984,eijtepjb1998}. Here we approximate the effective soft mode by a single Lorentz oscillator with an eigenvector given by the displacement of the atoms in the unit cell relative to their positions in the higher symmetry paraelectric phase. This mostly corresponds to a displacement of Sn$^{2+}$ ions along [100]. 

The induced atomic displacement $Q(z,t)$ along the soft mode eigenvector is then described by the equation of motion 
\begin{equation}
\ddot{Q}(z,t)+2 \Gamma \dot{Q}(z,t) +\omega_0^2 Q(z,t)=\beta E(z,t). \label{eq:EoM}
\end{equation}
The left side of the equation describes the lattice motion in a harmonic potential with a characteristic angular frequency $\omega_0$ and damping rate $\Gamma$. The right side of the equation describes the driving force which is equal to the applied electric field $E(z,t)$ in the slice multiplied by a scaling factor $\beta$. To estimate $E(z,t)$ from the incident field $E_i(t)$, we need to take into account the transmission coefficient and the absorption of the sample.  Given the complex, frequency-dependent index of refraction $\tilde{n}(\omega)$~\footnote{See  the supplemental information for a description of how the index of refraction is estimated self consistently within the framework of the fitting process.} and the incidence angle $\theta$ of the THz pulses with respect to the surface normal, we obtain in the frequency domain $E(z,\omega) = t_{\parallel}(\omega)\ e^{-\alpha_{T}(\omega) z}\ E_{i}(\omega)$, where $\alpha_T = 2 \omega k(\omega)/c$ and 
\begin{equation}
t_{\parallel}(\omega) = \frac{2\ n(\omega)\ \cos(\theta)}{n(\omega)^2\cos(\theta) + \sqrt{1-\sin^2(\theta)}}
\end{equation}
An inverse transform of $E(z,\omega)$ to the time domain yields $E(z,t)$.

The transient x-ray scattering factor $F(z,t)$ of a thin slice can then be determined from Eq.~\ref{eq:EoM}. To the first order it can be written as $F(z,t) = F(t_0) + \frac{\partial F}{\partial Q} Q(t) e^{-\alpha_{T}z}$. To finally calculate the normalized diffraction intensity, we integrate the structure factor over all slices to obtain the total scattering factor. The normalized diffraction intensity then becomes $\Delta I(t) / I_0 \propto \left| \int_0^{\infty} F (z,t) e^{-\alpha_x z } dz \right|^2$ where $\alpha_X$ is the absorption coefficient of the x-ray field. 

A least squares routine is used to fit the model to the experimental data by only adjusting the frequency, the damping rate and the coupling factor of the mode. A constant relative time shift is also needed in order to successfully fit the measured data. This time shift accounts for the uncertainty in the sample placement and was found to be less than 1 ps for all measurements. The results of the model fit to the experimental data are shown in Figure \ref{l.figure2} as a solid line for both measurements. We extract values of $f= 0.72 \pm 0.06$ THz for the frequency , $\Gamma= 0.3 \pm 0.1\ \mathrm{ps^{-1}}$ for the damping rate and $\beta = (1.8 \pm 0.5)\times10^{-6}\ \mathrm{C/\sqrt{kg}}$  for the coupling factor. The mode frequency and the damping rate compare well to parameters of the ferroelectric soft mode reported at 300 K \cite{vysochanskiiftt1978,vysochanskiivup2008}. The coupling factor $\beta= |e| \sum_{\eta,i} Z^*_{\eta,x i} \epsilon_{\eta,i}$~\footnote{See  the supplemental information at [URL] for a description of the proportionality factor.} can be estimated with the Born effective charge (BEC) $Z^*$ \cite{RushchanskiiPRL2007} and the mode eigenvector $\epsilon$ and yields a value of  $2.42 \times10^{-6}\ \mathrm{C/\sqrt{kg}}$, which is somewhat higher than the value obtained with the fit. We attribute this discrepancy mainly to an overestimation of the local fields, which we have assumed to be equivalent to the applied field. Taking these effects into account would require a calculation of the local dipole contribution to the electric field at each basis ion. 
\begin{figure}[h]
\includegraphics[width = 8.6cm]{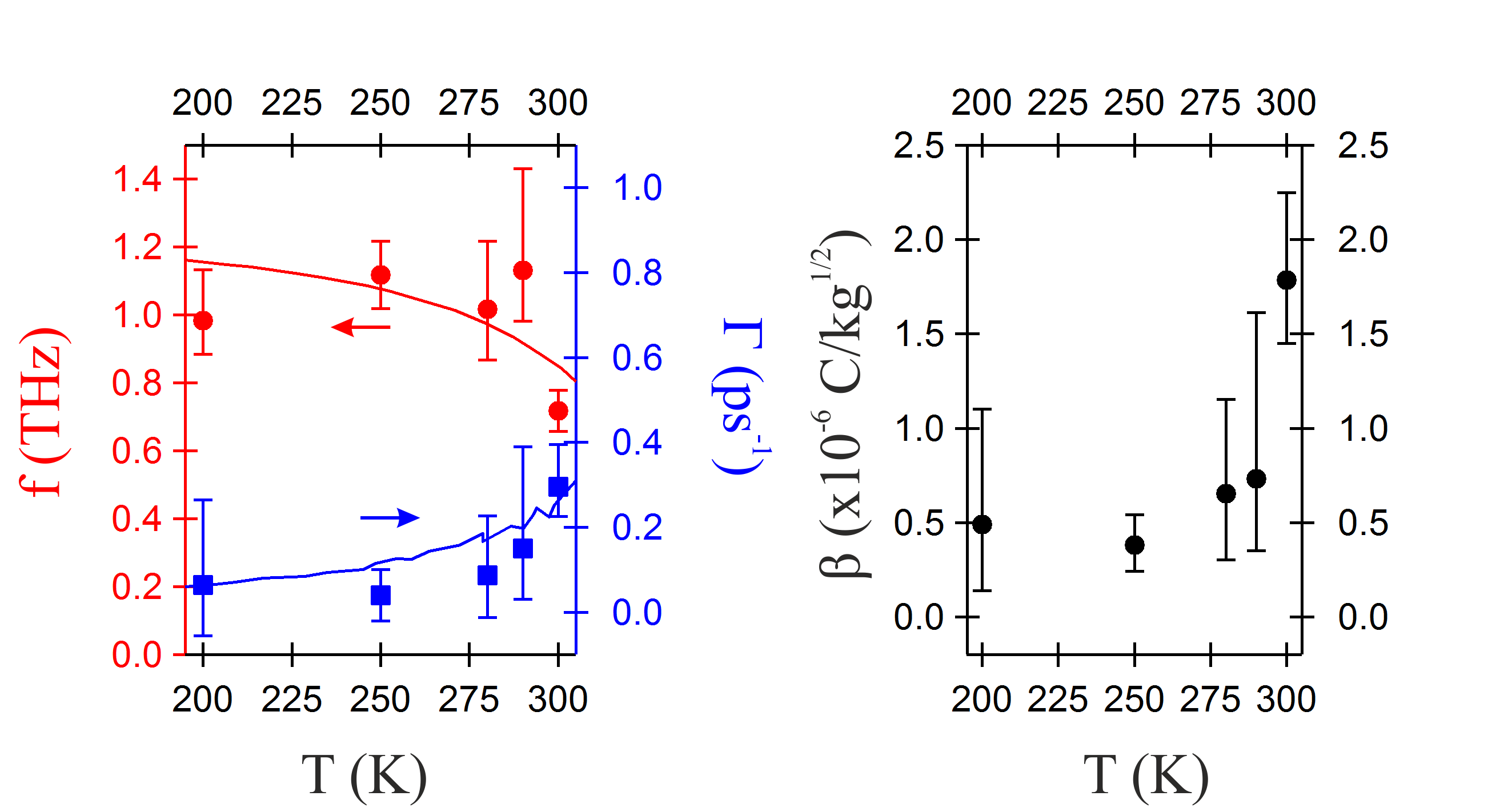}
\caption{\label{l.figure3}(full circles) Temperature dependence of the model fitting parameters: frequency $f$, damping rate $\Gamma$ and coupling coefficient $\beta$ (see text for a detailed description). (full lines) Experimental soft mode parameters extracted from \cite{vysochanskiiftt1978} and \cite{SlivkaFTT1979}}
\end{figure}

To measure the temperature dependence of the excited mode, additional data were taken at 290 K, 280 K, 250 K and 200 K. Figure \ref{l.figure3} shows the temperature dependence of the fitting parameters. We observe an increase of the frequency at lower temperatures and a decrease of the damping rate. A similar behavior was observed in Raman and neutron diffraction measurements \cite{eijtepjb1998,vysochanskiiftt1978,vysochanskiivup2008}. The coupling factor $\beta$ also decreases with temperature, mostly when cooling from 300 K to 290 K.

\begin{figure}
\includegraphics[width = 8.6cm]{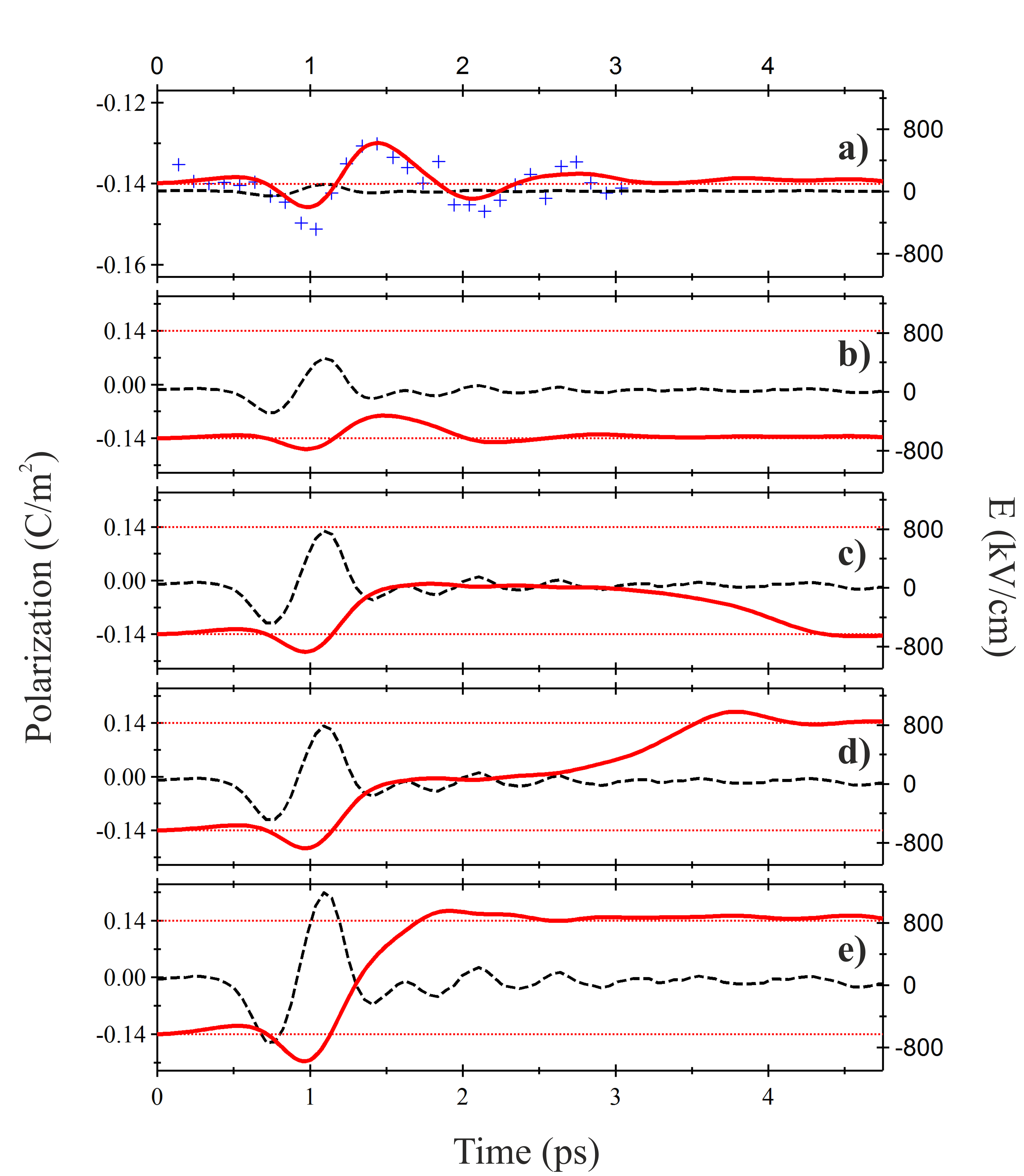}
\caption{\label{l.figure4} a) (symbol) Calculated polarization change of the measurement shown in Figure \ref{l.figure2}(a). The solid line is a fit to the data using an anharmonic model describing the ferroelectric potential (see text). The driving THz pulse (dashed line) and the calculated spontaneous ferroelectric state (dotted line) are also shown. b-e) Simulated polarization for increasing THz field strengths using potential parameters obtained in a).} 
\end{figure}

The maximum measured displacement of the two non-equivalent Sn$^{2+}$ ions are 2.3 pm and 1.5 pm. This corresponds to a 7.5\% displacement compared to the Sn$^{2+}$ ion displacement from the low to the high temperature phase. Using BECs and the atomic displacements, we calculate a 8.0\% polarization change compared to the spontaneous polarization magnitude of $13\ \mathrm{\mu C/cm^2}$ along the x direction \cite{CarpentierMRB1974} which is consistent with the obtained Sn$^{2+}$ ion displacement. The small displacement agrees well with the approximation of the ferroelectric ground state by a harmonic potential. For a displacement of the atoms further away from the ferroelectric ground state, the increasing anharmonicity of the potential would require higher order terms to describe the potential energy surface \cite{KatayamaPRL2012}. To estimate the field strength needed to switch the ferroelectric ground state, we expand our model by a quartic term to describe the energy potential along the soft mode coordinate as a double well potential and calculate the polarization at higher incident THz field amplitudes. Instead of equation (\ref{eq:EoM}), the equation of motion now reads $ \ddot{Q}(z,t)+2 \Gamma \dot{Q}(z,t) -a Q(z,t) + b Q(z,t)^3=\beta E(z,t)$ where $a$ and $b$ are given by the oscillation frequency and the distance between both ferroelectric states. The result is shown in Figure \ref{l.figure4}. At field strengths below 780 kV/cm (a-c) the ferroelectric polarization always relaxes to the same polarization state. At the threshold value of $\sim$ 790 kV/cm (d) the structure overcomes the potential wall and relaxes into the ground state of the opposite ferroelectric polarization. At higher field strength (e), the system relaxes faster to the opposite polarization state. We emphasize, however, that these results are based on a simple model that does not take into account the temperature dependence of the mode polarization and coupling to higher frequency modes.    

In conclusion, we have shown that the structural dynamics induced by single cycle THz pulses in a ferroelectric can clearly be resolved. Using ultrafast x-ray diffraction we directly observed a coherent motion of the atoms in response to the applied THz field. The temperature dependence of the frequency and the damping rate of the harmonic oscillation is a strong indication that the coherent motion corresponds to the ferroelectric soft mode of the material driven by the electric field of the THz pulses. The approximation of the soft mode by a single Lorentz oscillator and the polarization state by a damped harmonic potential is sufficient to describe the excitation process of the soft mode and to reproduce the transient diffraction changes. Our results suggest that increasing the amplitude of the terahertz electric field to approximately 1 MV/cm could lead to picosecond switching of the ferroelectric polarization.  
\begin{acknowledgments}
Time resolved x-ray diffraction measurements were carried out at the X05LA beam line of the Swiss Light Source, Paul Scherrer Institut, Villigen. We thank D. Grolimund and C. Borca for experimental help. We acknowledge financial support by the NCCR Molecular Ultrafast Science and Technology (NCCR MUST), a research instrument of the Swiss National Science Foundation (SNSF).
\end{acknowledgments}

\bibliographystyle{apsrev4-1}
\bibliography{library}

\begin{thebibliography}{31}%
\makeatletter
\providecommand \@ifxundefined [1]{%
 \@ifx{#1\undefined}
}%
\providecommand \@ifnum [1]{%
 \ifnum #1\expandafter \@firstoftwo
 \else \expandafter \@secondoftwo
 \fi
}%
\providecommand \@ifx [1]{%
 \ifx #1\expandafter \@firstoftwo
 \else \expandafter \@secondoftwo
 \fi
}%
\providecommand \natexlab [1]{#1}%
\providecommand \enquote  [1]{``#1''}%
\providecommand \bibnamefont  [1]{#1}%
\providecommand \bibfnamefont [1]{#1}%
\providecommand \citenamefont [1]{#1}%
\providecommand \href@noop [0]{\@secondoftwo}%
\providecommand \href [0]{\begingroup \@sanitize@url \@href}%
\providecommand \@href[1]{\@@startlink{#1}\@@href}%
\providecommand \@@href[1]{\endgroup#1\@@endlink}%
\providecommand \@sanitize@url [0]{\catcode `\\12\catcode `\$12\catcode
  `\&12\catcode `\#12\catcode `\^12\catcode `\_12\catcode `\%12\relax}%
\providecommand \@@startlink[1]{}%
\providecommand \@@endlink[0]{}%
\providecommand \url  [0]{\begingroup\@sanitize@url \@url }%
\providecommand \@url [1]{\endgroup\@href {#1}{\urlprefix }}%
\providecommand \urlprefix  [0]{URL }%
\providecommand \Eprint [0]{\href }%
\providecommand \doibase [0]{http://dx.doi.org/}%
\providecommand \selectlanguage [0]{\@gobble}%
\providecommand \bibinfo  [0]{\@secondoftwo}%
\providecommand \bibfield  [0]{\@secondoftwo}%
\providecommand \translation [1]{[#1]}%
\providecommand \BibitemOpen [0]{}%
\providecommand \bibitemStop [0]{}%
\providecommand \bibitemNoStop [0]{.\EOS\space}%
\providecommand \EOS [0]{\spacefactor3000\relax}%
\providecommand \BibitemShut  [1]{\csname bibitem#1\endcsname}%
\let\auto@bib@innerbib\@empty
\bibitem [{\citenamefont {Tybell}\ \emph {et~al.}(2002)\citenamefont {Tybell},
  \citenamefont {Paruch}, \citenamefont {Giamarchi},\ and\ \citenamefont
  {Triscone}}]{tybellprl89}%
  \BibitemOpen
  \bibfield  {author} {\bibinfo {author} {\bibfnamefont {T.}~\bibnamefont
  {Tybell}}, \bibinfo {author} {\bibfnamefont {P.}~\bibnamefont {Paruch}},
  \bibinfo {author} {\bibfnamefont {T.}~\bibnamefont {Giamarchi}}, \ and\
  \bibinfo {author} {\bibfnamefont {J.~M.}\ \bibnamefont {Triscone}},\
  }\href@noop {} {\bibfield  {journal} {\bibinfo  {journal} {Physical Review
  Letters}\ }\textbf {\bibinfo {volume} {89}},\ \bibinfo {pages} {097601}
  (\bibinfo {year} {2002})}\BibitemShut {NoStop}%
\bibitem [{\citenamefont {Grigoriev}\ \emph {et~al.}(2006)\citenamefont
  {Grigoriev}, \citenamefont {Do}, \citenamefont {Kim}, \citenamefont {Eom},
  \citenamefont {Adams}, \citenamefont {Dufresne},\ and\ \citenamefont
  {Evans}}]{grigorievprl2006}%
  \BibitemOpen
  \bibfield  {author} {\bibinfo {author} {\bibfnamefont {A.}~\bibnamefont
  {Grigoriev}}, \bibinfo {author} {\bibfnamefont {D.-H.}\ \bibnamefont {Do}},
  \bibinfo {author} {\bibfnamefont {D.~M.}\ \bibnamefont {Kim}}, \bibinfo
  {author} {\bibfnamefont {C.-B.}\ \bibnamefont {Eom}}, \bibinfo {author}
  {\bibfnamefont {B.}~\bibnamefont {Adams}}, \bibinfo {author} {\bibfnamefont
  {E.~M.}\ \bibnamefont {Dufresne}}, \ and\ \bibinfo {author} {\bibfnamefont
  {P.~G.}\ \bibnamefont {Evans}},\ }\href@noop {} {\bibfield  {journal}
  {\bibinfo  {journal} {Physical Review Letters}\ }\textbf {\bibinfo {volume}
  {96}},\ \bibinfo {pages} {187601} (\bibinfo {year} {2006})}\BibitemShut
  {NoStop}%
\bibitem [{\citenamefont {Fahy}\ and\ \citenamefont
  {Merlin}(1994)}]{fahyprl1994}%
  \BibitemOpen
  \bibfield  {author} {\bibinfo {author} {\bibfnamefont {S.}~\bibnamefont
  {Fahy}}\ and\ \bibinfo {author} {\bibfnamefont {R.}~\bibnamefont {Merlin}},\
  }\href@noop {} {\bibfield  {journal} {\bibinfo  {journal} {Physical Review
  Letters}\ }\textbf {\bibinfo {volume} {73}},\ \bibinfo {pages} {1122}
  (\bibinfo {year} {1994})}\BibitemShut {NoStop}%
\bibitem [{\citenamefont {Yan}\ and\ \citenamefont
  {Nelson}(1987)}]{yanjpc1987}%
  \BibitemOpen
  \bibfield  {author} {\bibinfo {author} {\bibfnamefont {Y.}~\bibnamefont
  {Yan}}\ and\ \bibinfo {author} {\bibfnamefont {K.~A.}\ \bibnamefont
  {Nelson}},\ }\href {\doibase doi:http://dx.doi.org/10.1063/1.453733}
  {\bibfield  {journal} {\bibinfo  {journal} {The Journal of Chemical Physics}\
  }\textbf {\bibinfo {volume} {87}},\ \bibinfo {pages} {6240} (\bibinfo {year}
  {1987})}\BibitemShut {NoStop}%
\bibitem [{\citenamefont {Zeiger}\ \emph {et~al.}(1992)\citenamefont {Zeiger},
  \citenamefont {Vidal}, \citenamefont {Cheng}, \citenamefont {Ippen},
  \citenamefont {Dresselhaus},\ and\ \citenamefont
  {Dresselhaus}}]{zeigerprb1992}%
  \BibitemOpen
  \bibfield  {author} {\bibinfo {author} {\bibfnamefont {H.~J.}\ \bibnamefont
  {Zeiger}}, \bibinfo {author} {\bibfnamefont {J.}~\bibnamefont {Vidal}},
  \bibinfo {author} {\bibfnamefont {T.~K.}\ \bibnamefont {Cheng}}, \bibinfo
  {author} {\bibfnamefont {E.~P.}\ \bibnamefont {Ippen}}, \bibinfo {author}
  {\bibfnamefont {G.}~\bibnamefont {Dresselhaus}}, \ and\ \bibinfo {author}
  {\bibfnamefont {M.~S.}\ \bibnamefont {Dresselhaus}},\ }\href {\doibase
  10.1103/PhysRevB.45.768} {\bibfield  {journal} {\bibinfo  {journal} {Phys.
  Rev. B}\ }\textbf {\bibinfo {volume} {45}},\ \bibinfo {pages} {768} (\bibinfo
  {year} {1992})}\BibitemShut {NoStop}%
\bibitem [{\citenamefont {Qi}\ \emph {et~al.}(2009)\citenamefont {Qi},
  \citenamefont {Shin}, \citenamefont {Yeh}, \citenamefont {Nelson},\ and\
  \citenamefont {Rappe}}]{qiprl2009}%
  \BibitemOpen
  \bibfield  {author} {\bibinfo {author} {\bibfnamefont {T.}~\bibnamefont
  {Qi}}, \bibinfo {author} {\bibfnamefont {Y.-H.}\ \bibnamefont {Shin}},
  \bibinfo {author} {\bibfnamefont {K.-L.}\ \bibnamefont {Yeh}}, \bibinfo
  {author} {\bibfnamefont {K.~A.}\ \bibnamefont {Nelson}}, \ and\ \bibinfo
  {author} {\bibfnamefont {A.~M.}\ \bibnamefont {Rappe}},\ }\href@noop {}
  {\bibfield  {journal} {\bibinfo  {journal} {Physical Review Letters}\
  }\textbf {\bibinfo {volume} {102}},\ \bibinfo {pages} {247603} (\bibinfo
  {year} {2009})}\BibitemShut {NoStop}%
\bibitem [{\citenamefont {Hirori}\ \emph {et~al.}(2011)\citenamefont {Hirori},
  \citenamefont {Doi}, \citenamefont {Blanchard},\ and\ \citenamefont
  {Tanaka}}]{HiroriAPL2011}%
  \BibitemOpen
  \bibfield  {author} {\bibinfo {author} {\bibfnamefont {H.}~\bibnamefont
  {Hirori}}, \bibinfo {author} {\bibfnamefont {A.}~\bibnamefont {Doi}},
  \bibinfo {author} {\bibfnamefont {F.}~\bibnamefont {Blanchard}}, \ and\
  \bibinfo {author} {\bibfnamefont {K.}~\bibnamefont {Tanaka}},\ }\href
  {\doibase doi:http://dx.doi.org/10.1063/1.3560062} {\bibfield  {journal}
  {\bibinfo  {journal} {Applied Physics Letters}\ }\textbf {\bibinfo {volume}
  {98}},\ \bibinfo {pages} {091106} (\bibinfo {year} {2011})}\BibitemShut
  {NoStop}%
\bibitem [{\citenamefont {Vicario}\ \emph {et~al.}(2013)\citenamefont
  {Vicario}, \citenamefont {Ruchert},\ and\ \citenamefont
  {Hauri}}]{VicarioJMO2013}%
  \BibitemOpen
  \bibfield  {author} {\bibinfo {author} {\bibfnamefont {C.}~\bibnamefont
  {Vicario}}, \bibinfo {author} {\bibfnamefont {C.}~\bibnamefont {Ruchert}}, \
  and\ \bibinfo {author} {\bibfnamefont {C.~P.}\ \bibnamefont {Hauri}},\ }\href
  {\doibase 10.1080/09500340.2013.800242} {\bibfield  {journal} {\bibinfo
  {journal} {Journal of Modern Optics}\ ,\ \bibinfo {pages} {1}} (\bibinfo
  {year} {2013})}\BibitemShut {NoStop}%
\bibitem [{\citenamefont {Shalaby}\ and\ \citenamefont
  {Hauri}(2015)}]{ShalabyNC2015}%
  \BibitemOpen
  \bibfield  {author} {\bibinfo {author} {\bibfnamefont {M.}~\bibnamefont
  {Shalaby}}\ and\ \bibinfo {author} {\bibfnamefont {C.~P.}\ \bibnamefont
  {Hauri}},\ }\href@noop {} {\bibfield  {journal} {\bibinfo  {journal} {Nature
  Communications}\ }\textbf {\bibinfo {volume} {6}} (\bibinfo {year}
  {2015})}\BibitemShut {NoStop}%
\bibitem [{\citenamefont {Katayama}\ \emph {et~al.}(2012)\citenamefont
  {Katayama}, \citenamefont {Aoki}, \citenamefont {Takeda}, \citenamefont
  {Shimosato}, \citenamefont {Ashida}, \citenamefont {Kinjo}, \citenamefont
  {Kawayama}, \citenamefont {Tonouchi}, \citenamefont {Nagai},\ and\
  \citenamefont {Tanaka}}]{KatayamaPRL2012}%
  \BibitemOpen
  \bibfield  {author} {\bibinfo {author} {\bibfnamefont {I.}~\bibnamefont
  {Katayama}}, \bibinfo {author} {\bibfnamefont {H.}~\bibnamefont {Aoki}},
  \bibinfo {author} {\bibfnamefont {J.}~\bibnamefont {Takeda}}, \bibinfo
  {author} {\bibfnamefont {H.}~\bibnamefont {Shimosato}}, \bibinfo {author}
  {\bibfnamefont {M.}~\bibnamefont {Ashida}}, \bibinfo {author} {\bibfnamefont
  {R.}~\bibnamefont {Kinjo}}, \bibinfo {author} {\bibfnamefont
  {I.}~\bibnamefont {Kawayama}}, \bibinfo {author} {\bibfnamefont
  {M.}~\bibnamefont {Tonouchi}}, \bibinfo {author} {\bibfnamefont
  {M.}~\bibnamefont {Nagai}}, \ and\ \bibinfo {author} {\bibfnamefont
  {K.}~\bibnamefont {Tanaka}},\ }\href@noop {} {\bibfield  {journal} {\bibinfo
  {journal} {Physical Review Letters}\ }\textbf {\bibinfo {volume} {108}},\
  \bibinfo {pages} {097401} (\bibinfo {year} {2012})}\BibitemShut {NoStop}%
\bibitem [{\citenamefont {Els\"asser}\ and\ \citenamefont
  {W\"orner}(2014)}]{ElsaesserJCP2014}%
  \BibitemOpen
  \bibfield  {author} {\bibinfo {author} {\bibfnamefont {T.}~\bibnamefont
  {Els\"asser}}\ and\ \bibinfo {author} {\bibfnamefont {M.}~\bibnamefont
  {W\"orner}},\ }\href {\doibase doi:http://dx.doi.org/10.1063/1.4855115}
  {\bibfield  {journal} {\bibinfo  {journal} {The Journal of Chemical Physics}\
  }\textbf {\bibinfo {volume} {140}},\ \bibinfo {pages} {020901} (\bibinfo
  {year} {2014})}\BibitemShut {NoStop}%
\bibitem [{\citenamefont {F\"orst}\ \emph {et~al.}(2015)\citenamefont
  {F\"orst}, \citenamefont {Mankowsky},\ and\ \citenamefont
  {Cavalleri}}]{FoerstACR2015}%
  \BibitemOpen
  \bibfield  {author} {\bibinfo {author} {\bibfnamefont {M.}~\bibnamefont
  {F\"orst}}, \bibinfo {author} {\bibfnamefont {R.}~\bibnamefont {Mankowsky}},
  \ and\ \bibinfo {author} {\bibfnamefont {A.}~\bibnamefont {Cavalleri}},\
  }\href {\doibase 10.1021/ar500391x} {\bibfield  {journal} {\bibinfo
  {journal} {Accounts of Chemical Research}\ }\textbf {\bibinfo {volume}
  {48}},\ \bibinfo {pages} {380} (\bibinfo {year} {2015})}\BibitemShut
  {NoStop}%
\bibitem [{\citenamefont {Cavalleri}\ \emph {et~al.}(2006)\citenamefont
  {Cavalleri}, \citenamefont {Wall}, \citenamefont {Simpson}, \citenamefont
  {Statz}, \citenamefont {Ward}, \citenamefont {Nelson}, \citenamefont {Rini},\
  and\ \citenamefont {Sch\"onlein}}]{CavalleriNature2006}%
  \BibitemOpen
  \bibfield  {author} {\bibinfo {author} {\bibfnamefont {A.}~\bibnamefont
  {Cavalleri}}, \bibinfo {author} {\bibfnamefont {S.}~\bibnamefont {Wall}},
  \bibinfo {author} {\bibfnamefont {C.}~\bibnamefont {Simpson}}, \bibinfo
  {author} {\bibfnamefont {E.}~\bibnamefont {Statz}}, \bibinfo {author}
  {\bibfnamefont {D.~W.}\ \bibnamefont {Ward}}, \bibinfo {author}
  {\bibfnamefont {K.~A.}\ \bibnamefont {Nelson}}, \bibinfo {author}
  {\bibfnamefont {M.}~\bibnamefont {Rini}}, \ and\ \bibinfo {author}
  {\bibfnamefont {R.~W.}\ \bibnamefont {Sch\"onlein}},\ }\href@noop {}
  {\bibfield  {journal} {\bibinfo  {journal} {Nature}\ }\textbf {\bibinfo
  {volume} {442}},\ \bibinfo {pages} {664} (\bibinfo {year}
  {2006})}\BibitemShut {NoStop}%
\bibitem [{\citenamefont {Dittmar}\ and\ \citenamefont
  {Sch\"afer}(1974)}]{dittmarznatur1974}%
  \BibitemOpen
  \bibfield  {author} {\bibinfo {author} {\bibfnamefont {G.}~\bibnamefont
  {Dittmar}}\ and\ \bibinfo {author} {\bibfnamefont {H.}~\bibnamefont
  {Sch\"afer}},\ }\href@noop {} {\bibfield  {journal} {\bibinfo  {journal} {Z.
  Naturfosch}\ }\textbf {\bibinfo {volume} {29b}},\ \bibinfo {pages} {312}
  (\bibinfo {year} {1974})}\BibitemShut {NoStop}%
\bibitem [{\citenamefont {Scott}\ \emph {et~al.}(1992)\citenamefont {Scott},
  \citenamefont {Pressprich}, \citenamefont {Willet},\ and\ \citenamefont
  {Cleary}}]{scottjssc1992}%
  \BibitemOpen
  \bibfield  {author} {\bibinfo {author} {\bibfnamefont {B.}~\bibnamefont
  {Scott}}, \bibinfo {author} {\bibfnamefont {M.}~\bibnamefont {Pressprich}},
  \bibinfo {author} {\bibfnamefont {R.~D.}\ \bibnamefont {Willet}}, \ and\
  \bibinfo {author} {\bibfnamefont {D.~A.}\ \bibnamefont {Cleary}},\ }\href
  {\doibase http://dx.doi.org/10.1016/S0022-4596(05)80262-2} {\bibfield
  {journal} {\bibinfo  {journal} {Journal of Solid State Chemistry}\ }\textbf
  {\bibinfo {volume} {96}},\ \bibinfo {pages} {294} (\bibinfo {year}
  {1992})}\BibitemShut {NoStop}%
\bibitem [{\citenamefont {Vysochanskii}\ \emph {et~al.}(1978)\citenamefont
  {Vysochanskii}, \citenamefont {Slivka}, \citenamefont {Buturlakin},
  \citenamefont {Gurzan},\ and\ \citenamefont {Chepur}}]{vysochanskiiftt1978}%
  \BibitemOpen
  \bibfield  {author} {\bibinfo {author} {\bibfnamefont {Y.~M.}\ \bibnamefont
  {Vysochanskii}}, \bibinfo {author} {\bibfnamefont {V.~Y.}\ \bibnamefont
  {Slivka}}, \bibinfo {author} {\bibfnamefont {A.~P.}\ \bibnamefont
  {Buturlakin}}, \bibinfo {author} {\bibfnamefont {M.~I.}\ \bibnamefont
  {Gurzan}}, \ and\ \bibinfo {author} {\bibfnamefont {D.~V.}\ \bibnamefont
  {Chepur}},\ }\href@noop {} {\bibfield  {journal} {\bibinfo  {journal} {Fiz.
  Tv. Tela}\ }\textbf {\bibinfo {volume} {20}} (\bibinfo {year}
  {1978})}\BibitemShut {NoStop}%
\bibitem [{\citenamefont {Grabar}\ \emph {et~al.}(1984)\citenamefont {Grabar},
  \citenamefont {Vysochanskii},\ and\ \citenamefont {Slivka}}]{grabarftt1984}%
  \BibitemOpen
  \bibfield  {author} {\bibinfo {author} {\bibfnamefont {A.~A.}\ \bibnamefont
  {Grabar}}, \bibinfo {author} {\bibfnamefont {Y.~M.}\ \bibnamefont
  {Vysochanskii}}, \ and\ \bibinfo {author} {\bibfnamefont {V.~Y.}\
  \bibnamefont {Slivka}},\ }\href@noop {} {\bibfield  {journal} {\bibinfo
  {journal} {Fiz. Tv. Tela}\ }\textbf {\bibinfo {volume} {26}} (\bibinfo {year}
  {1984})}\BibitemShut {NoStop}%
\bibitem [{\citenamefont {Eijt}\ \emph {et~al.}(1998)\citenamefont {Eijt},
  \citenamefont {Currat}, \citenamefont {Lorenzo}, \citenamefont
  {Saint-Gr\'egoire}, \citenamefont {Hennion},\ and\ \citenamefont
  {Vysochanskii}}]{eijtepjb1998}%
  \BibitemOpen
  \bibfield  {author} {\bibinfo {author} {\bibfnamefont {S.~W.}\ \bibnamefont
  {Eijt}}, \bibinfo {author} {\bibfnamefont {R.}~\bibnamefont {Currat}},
  \bibinfo {author} {\bibfnamefont {J.~E.}\ \bibnamefont {Lorenzo}}, \bibinfo
  {author} {\bibfnamefont {P.}~\bibnamefont {Saint-Gr\'egoire}}, \bibinfo
  {author} {\bibfnamefont {B.}~\bibnamefont {Hennion}}, \ and\ \bibinfo
  {author} {\bibfnamefont {Y.~M.}\ \bibnamefont {Vysochanskii}},\ }\href@noop
  {} {\bibfield  {journal} {\bibinfo  {journal} {Eur. Phys. J. B}\ }\textbf
  {\bibinfo {volume} {5}},\ \bibinfo {pages} {169} (\bibinfo {year}
  {1998})}\BibitemShut {NoStop}%
\bibitem [{\citenamefont {Beaud}\ \emph {et~al.}(2007)\citenamefont {Beaud},
  \citenamefont {Johnson}, \citenamefont {Streun}, \citenamefont {Abela},
  \citenamefont {Abramsohn}, \citenamefont {Grolimund}, \citenamefont
  {Krasniqi}, \citenamefont {Schmidt}, \citenamefont {Schlott},\ and\
  \citenamefont {Ingold}}]{beaudprl2007}%
  \BibitemOpen
  \bibfield  {author} {\bibinfo {author} {\bibfnamefont {P.}~\bibnamefont
  {Beaud}}, \bibinfo {author} {\bibfnamefont {S.~L.}\ \bibnamefont {Johnson}},
  \bibinfo {author} {\bibfnamefont {A.}~\bibnamefont {Streun}}, \bibinfo
  {author} {\bibfnamefont {R.}~\bibnamefont {Abela}}, \bibinfo {author}
  {\bibfnamefont {D.}~\bibnamefont {Abramsohn}}, \bibinfo {author}
  {\bibfnamefont {D.}~\bibnamefont {Grolimund}}, \bibinfo {author}
  {\bibfnamefont {F.}~\bibnamefont {Krasniqi}}, \bibinfo {author}
  {\bibfnamefont {T.}~\bibnamefont {Schmidt}}, \bibinfo {author} {\bibfnamefont
  {V.}~\bibnamefont {Schlott}}, \ and\ \bibinfo {author} {\bibfnamefont
  {G.}~\bibnamefont {Ingold}},\ }\href@noop {} {\bibfield  {journal} {\bibinfo
  {journal} {Physical Review Letters}\ }\textbf {\bibinfo {volume} {99}},\
  \bibinfo {pages} {174801} (\bibinfo {year} {2007})}\BibitemShut {NoStop}%
\bibitem [{\citenamefont {Brunner}\ \emph {et~al.}(2008)\citenamefont
  {Brunner}, \citenamefont {Kwon}, \citenamefont {Kwon}, \citenamefont
  {Jazbinsek}, \citenamefont {Schneider},\ and\ \citenamefont
  {G\"unter}}]{BrunnerOE2008}%
  \BibitemOpen
  \bibfield  {author} {\bibinfo {author} {\bibfnamefont {F.~D.}\ \bibnamefont
  {Brunner}}, \bibinfo {author} {\bibfnamefont {O.~P.}\ \bibnamefont {Kwon}},
  \bibinfo {author} {\bibfnamefont {S.-J.}\ \bibnamefont {Kwon}}, \bibinfo
  {author} {\bibfnamefont {M.}~\bibnamefont {Jazbinsek}}, \bibinfo {author}
  {\bibfnamefont {A.}~\bibnamefont {Schneider}}, \ and\ \bibinfo {author}
  {\bibfnamefont {P.}~\bibnamefont {G\"unter}},\ }\href {\doibase
  10.1364/OE.16.016496} {\bibfield  {journal} {\bibinfo  {journal} {Optics
  Express}\ }\textbf {\bibinfo {volume} {16}},\ \bibinfo {pages} {16496}
  (\bibinfo {year} {2008})}\BibitemShut {NoStop}%
\bibitem [{\citenamefont {Hunziker}\ \emph {et~al.}(2008)\citenamefont
  {Hunziker}, \citenamefont {Kwon}, \citenamefont {Figi}, \citenamefont
  {Juvalta}, \citenamefont {Kwon}, \citenamefont {Jazbinsek},\ and\
  \citenamefont {G\"unter}}]{HunzikerJOSB2008}%
  \BibitemOpen
  \bibfield  {author} {\bibinfo {author} {\bibfnamefont {C.}~\bibnamefont
  {Hunziker}}, \bibinfo {author} {\bibfnamefont {S.-J.}\ \bibnamefont {Kwon}},
  \bibinfo {author} {\bibfnamefont {H.}~\bibnamefont {Figi}}, \bibinfo {author}
  {\bibfnamefont {F.}~\bibnamefont {Juvalta}}, \bibinfo {author} {\bibfnamefont
  {O.~P.}\ \bibnamefont {Kwon}}, \bibinfo {author} {\bibfnamefont
  {M.}~\bibnamefont {Jazbinsek}}, \ and\ \bibinfo {author} {\bibfnamefont
  {P.}~\bibnamefont {G\"unter}},\ }\href {\doibase 10.1364/JOSAB.25.001678}
  {\bibfield  {journal} {\bibinfo  {journal} {Journal of the Optical Society of
  America B}\ }\textbf {\bibinfo {volume} {25}},\ \bibinfo {pages} {1678}
  (\bibinfo {year} {2008})}\BibitemShut {NoStop}%
\bibitem [{\citenamefont {Kwon}\ \emph {et~al.}(2008)\citenamefont {Kwon},
  \citenamefont {Kwon}, \citenamefont {Jazbinsek}, \citenamefont {Brunner},
  \citenamefont {Seo}, \citenamefont {Hunziker}, \citenamefont {Schneider},
  \citenamefont {Yun}, \citenamefont {Lee},\ and\ \citenamefont
  {G\"unter}}]{KwonAFM2008}%
  \BibitemOpen
  \bibfield  {author} {\bibinfo {author} {\bibfnamefont {O.~P.}\ \bibnamefont
  {Kwon}}, \bibinfo {author} {\bibfnamefont {S.-J.}\ \bibnamefont {Kwon}},
  \bibinfo {author} {\bibfnamefont {M.}~\bibnamefont {Jazbinsek}}, \bibinfo
  {author} {\bibfnamefont {F.~D.~J.}\ \bibnamefont {Brunner}}, \bibinfo
  {author} {\bibfnamefont {J.-I.}\ \bibnamefont {Seo}}, \bibinfo {author}
  {\bibfnamefont {C.}~\bibnamefont {Hunziker}}, \bibinfo {author}
  {\bibfnamefont {A.}~\bibnamefont {Schneider}}, \bibinfo {author}
  {\bibfnamefont {H.}~\bibnamefont {Yun}}, \bibinfo {author} {\bibfnamefont
  {Y.-S.}\ \bibnamefont {Lee}}, \ and\ \bibinfo {author} {\bibfnamefont
  {P.}~\bibnamefont {G\"unter}},\ }\href {\doibase 10.1002/adfm.200800633}
  {\bibfield  {journal} {\bibinfo  {journal} {Advanced Functional Materials}\
  }\textbf {\bibinfo {volume} {18}},\ \bibinfo {pages} {3242} (\bibinfo {year}
  {2008})}\BibitemShut {NoStop}%
\bibitem [{\citenamefont {Ruchert}\ \emph {et~al.}(2012)\citenamefont
  {Ruchert}, \citenamefont {Vicario},\ and\ \citenamefont
  {Hauri}}]{RuchertOL2012}%
  \BibitemOpen
  \bibfield  {author} {\bibinfo {author} {\bibfnamefont {C.}~\bibnamefont
  {Ruchert}}, \bibinfo {author} {\bibfnamefont {C.}~\bibnamefont {Vicario}}, \
  and\ \bibinfo {author} {\bibfnamefont {C.~P.}\ \bibnamefont {Hauri}},\ }\href
  {\doibase 10.1364/OL.37.000899} {\bibfield  {journal} {\bibinfo  {journal}
  {Optics Letters}\ }\textbf {\bibinfo {volume} {37}},\ \bibinfo {pages} {899}
  (\bibinfo {year} {2012})}\BibitemShut {NoStop}%
\bibitem [{\citenamefont {Johnson}\ \emph {et~al.}(2008)\citenamefont
  {Johnson}, \citenamefont {Beaud}, \citenamefont {Milne}, \citenamefont
  {Krasniqi}, \citenamefont {Zijlstra}, \citenamefont {Garcia}, \citenamefont
  {Kaiser}, \citenamefont {Grolimund}, \citenamefont {Abela},\ and\
  \citenamefont {Ingold}}]{JohnsonPRL2008}%
  \BibitemOpen
  \bibfield  {author} {\bibinfo {author} {\bibfnamefont {S.~L.}\ \bibnamefont
  {Johnson}}, \bibinfo {author} {\bibfnamefont {P.}~\bibnamefont {Beaud}},
  \bibinfo {author} {\bibfnamefont {C.~J.}\ \bibnamefont {Milne}}, \bibinfo
  {author} {\bibfnamefont {F.~S.}\ \bibnamefont {Krasniqi}}, \bibinfo {author}
  {\bibfnamefont {E.~S.}\ \bibnamefont {Zijlstra}}, \bibinfo {author}
  {\bibfnamefont {M.~E.}\ \bibnamefont {Garcia}}, \bibinfo {author}
  {\bibfnamefont {M.}~\bibnamefont {Kaiser}}, \bibinfo {author} {\bibfnamefont
  {D.}~\bibnamefont {Grolimund}}, \bibinfo {author} {\bibfnamefont
  {R.}~\bibnamefont {Abela}}, \ and\ \bibinfo {author} {\bibfnamefont
  {G.}~\bibnamefont {Ingold}},\ }\href@noop {} {\bibfield  {journal} {\bibinfo
  {journal} {Physical Review Letters}\ }\textbf {\bibinfo {volume} {100}},\
  \bibinfo {pages} {155501} (\bibinfo {year} {2008})}\BibitemShut {NoStop}%
\bibitem [{\citenamefont {Volkov}\ \emph {et~al.}(1983)\citenamefont {Volkov},
  \citenamefont {Kozlov}, \citenamefont {Afanaseva}, \citenamefont {Yu},
  \citenamefont {Vysochanskii}, \citenamefont {Grabar},\ and\ \citenamefont
  {Slivka}}]{volkovSPSS1983}%
  \BibitemOpen
  \bibfield  {author} {\bibinfo {author} {\bibfnamefont {A.}~\bibnamefont
  {Volkov}}, \bibinfo {author} {\bibfnamefont {G.}~\bibnamefont {Kozlov}},
  \bibinfo {author} {\bibfnamefont {N.}~\bibnamefont {Afanaseva}}, \bibinfo
  {author} {\bibfnamefont {M.}~\bibnamefont {Yu}}, \bibinfo {author}
  {\bibfnamefont {Y.~M.}\ \bibnamefont {Vysochanskii}}, \bibinfo {author}
  {\bibfnamefont {A.~A.}\ \bibnamefont {Grabar}}, \ and\ \bibinfo {author}
  {\bibfnamefont {V.~Y.}\ \bibnamefont {Slivka}},\ }\href@noop {} {\bibfield
  {journal} {\bibinfo  {journal} {Sov. Phys. Solid State}\ }\textbf {\bibinfo
  {volume} {25}} (\bibinfo {year} {1983})}\BibitemShut {NoStop}%
\bibitem [{Note1()}]{Note1}%
  \BibitemOpen
  \bibinfo {note} {See the supplemental information for a description of how
  the index of refraction is estimated self consistently within the framework
  of the fitting process.}\BibitemShut {Stop}%
\bibitem [{\citenamefont {Vysochanskii}\ \emph {et~al.}(2008)\citenamefont
  {Vysochanskii}, \citenamefont {Janssen}, \citenamefont {Currat},
  \citenamefont {Folk}, \citenamefont {Banys}, \citenamefont {Grigas},\ and\
  \citenamefont {Samulionis}}]{vysochanskiivup2008}%
  \BibitemOpen
  \bibfield  {author} {\bibinfo {author} {\bibfnamefont {Y.~M.}\ \bibnamefont
  {Vysochanskii}}, \bibinfo {author} {\bibfnamefont {T.}~\bibnamefont
  {Janssen}}, \bibinfo {author} {\bibfnamefont {R.}~\bibnamefont {Currat}},
  \bibinfo {author} {\bibfnamefont {R.}~\bibnamefont {Folk}}, \bibinfo {author}
  {\bibfnamefont {J.}~\bibnamefont {Banys}}, \bibinfo {author} {\bibfnamefont
  {J.}~\bibnamefont {Grigas}}, \ and\ \bibinfo {author} {\bibfnamefont
  {V.}~\bibnamefont {Samulionis}},\ }\href@noop {} {\emph {\bibinfo {title}
  {Phase transitions in ferroelectric phosphorus chalcogenide crystals}}},\
  \bibinfo {edition} {2nd}\ ed.\ (\bibinfo  {publisher} {Vilnius University
  Publishing House, Lithuania},\ \bibinfo {address} {Vilnius, Lithuania},\
  \bibinfo {year} {2008})\ p.\ \bibinfo {pages} {455}\BibitemShut {NoStop}%
\bibitem [{Note2()}]{Note2}%
  \BibitemOpen
  \bibinfo {note} {See the supplemental information at [URL] for a description
  of the proportionality factor.}\BibitemShut {Stop}%
\bibitem [{\citenamefont {Rushchanskii}\ \emph {et~al.}(2007)\citenamefont
  {Rushchanskii}, \citenamefont {Vysochanskii},\ and\ \citenamefont
  {Strauch}}]{RushchanskiiPRL2007}%
  \BibitemOpen
  \bibfield  {author} {\bibinfo {author} {\bibfnamefont {K.~Z.}\ \bibnamefont
  {Rushchanskii}}, \bibinfo {author} {\bibfnamefont {Y.~M.}\ \bibnamefont
  {Vysochanskii}}, \ and\ \bibinfo {author} {\bibfnamefont {D.}~\bibnamefont
  {Strauch}},\ }\href@noop {} {\bibfield  {journal} {\bibinfo  {journal}
  {Physical Review Letters}\ }\textbf {\bibinfo {volume} {99}},\ \bibinfo
  {pages} {207601} (\bibinfo {year} {2007})}\BibitemShut {NoStop}%
\bibitem [{\citenamefont {Slivka}\ \emph {et~al.}(1979)\citenamefont {Slivka},
  \citenamefont {Vysochanskii}, \citenamefont {Gurzan},\ and\ \citenamefont
  {Chepur}}]{SlivkaFTT1979}%
  \BibitemOpen
  \bibfield  {author} {\bibinfo {author} {\bibfnamefont {V.~Y.}\ \bibnamefont
  {Slivka}}, \bibinfo {author} {\bibfnamefont {Y.~M.}\ \bibnamefont
  {Vysochanskii}}, \bibinfo {author} {\bibfnamefont {M.~I.}\ \bibnamefont
  {Gurzan}}, \ and\ \bibinfo {author} {\bibfnamefont {D.~V.}\ \bibnamefont
  {Chepur}},\ }\href@noop {} {\bibfield  {journal} {\bibinfo  {journal} {Fiz.
  Tv. Tela}\ }\textbf {\bibinfo {volume} {21}},\ \bibinfo {pages} {2396}
  (\bibinfo {year} {1979})}\BibitemShut {NoStop}%
\bibitem [{\citenamefont {Carpentier}\ and\ \citenamefont
  {Nitsche}(1974)}]{CarpentierMRB1974}%
  \BibitemOpen
  \bibfield  {author} {\bibinfo {author} {\bibfnamefont {C.~D.}\ \bibnamefont
  {Carpentier}}\ and\ \bibinfo {author} {\bibfnamefont {R.}~\bibnamefont
  {Nitsche}},\ }\href {\doibase http://dx.doi.org/10.1016/0025-5408(74)90023-3}
  {\bibfield  {journal} {\bibinfo  {journal} {Materials Research Bulletin}\
  }\textbf {\bibinfo {volume} {9}},\ \bibinfo {pages} {1097} (\bibinfo {year}
  {1974})}\BibitemShut {NoStop}%
\end{thebibliography}%

\end{document}